# Correlations Between Switching of Conductivity and Optical Radiation Observed in Thin Graphite-Like Films

S.G.Lebedev[1] V.E.Yants[1] and A.S.Lebedev[2]

[1]Institute for Nuclear Research of Russian Academy of Sciences
60[th] October Anniversary Prospect, 7a
117312, Moscow, Russia
[2]Lomonosov Moscow State University,
Faculty of Computational Mathematics and Cybernetics
GSP-2, Vorobievy Gory, 119992, Moscow, Russia

The satisfactory explanation of abnormal electromagnetics in thin graphite-like carbon films till now is absent. The most comprehensible explanation may be the high-temperature superconductivity (*HTSC*). Times of spasmodic switching of electrical conductivity are measured in this work in graphite-like nanostructured carbon films, produced by methods of the carbon arc (*CA*) and chemical vapor deposition (*CVD*), which have made *1* and *2* nanoseconds correspondingly. So fast switching completely excludes the thermal mechanism of the process. According to *HTSC* logic, in the time vicinity close to jump of electroresistance it is necessary to expect the generation of optical radiation in infrared (*IR*) range. This work presents the first results on registration of *IR* radiation caused by sharp change of conductivity in thin graphite-like carbon films.

## 1. Introduction

At the study of conductive properties of some carbon condensates the phenomenon of spasmodic increase in electrical resistance on *~4-5* orders of magnitude under electric current increase up to some critical value is revealed. The critical current decreases with temperature and at a room temperature has the values of *5 - 500 mA* (depending on deposition condition of a carbon film and its further processing) at the applied DC voltage in the range of *5-50 V*. Experimental samples represented a carbon film with the thickness of *1000-2000* Angstrom, with the sizes in plane of *1? 0.5 $cm^2$* though these sizes can be considerably reduced without appreciable influence on the results of experiment [1-2]. More than *30* years ago such kind of behavior was observed in glassy carbon [3] and carbon deposits [4] by famous polish scientist Kazimierz Antonowicz (1914-2002). It is turning out that electroresistance of these systems at some threshold voltage (*2-10 V*) spasmodically drops approximately on an order of magnitude. As the conditions of producing of carbon films in [1] are rather close to those [3-4], it is possible to assume a similarity of switching effects in these two cases. However here are also essential differences. At first, the jump of resistance in our experimnts makes *4-5* orders of its magnitude (see Fig.1). Secondly, the jump of resistance in [1] in difference from [3-4] regularly depends on temperature. Later on K.Antonowicz produce the *Al-C-Al* sandvich with thin *10*-Angstrom layer of amorphous carbon in which he observed the resistance jump up to *3* orders of magnitude and the effect of microwave detection. The later manifest oneself by the influence of microvawe irradiation on the transport current and were explained due to Josephson effect in single Josephson contact at room temperature [5]. The resistance of sample in Antonowicz's experiments was recovered after relaxation time of about few days. In our experiments we also observed the relaxation times of the same order of magnitude. The reason for such relaxation is not clear so far however we have some idea about its origin, which is listed below. During long time we have not been awared on Antonowicz's results and reproduced many of his findings independently but on the base of principly another statement and fashion as it will be seen below.

## 2. Exploration of Josephson media in graphite-like carbon films.

Unlike Antonowicz we proceed from the fact that carbon film composed by small graphite-like granules embedded in the matrix of amorphous carbon. Each granule of that structure acts as josphson contact. So the structure of carbon film studied is the kind of Josephson media as it is finded in [6]. Such kind of system is ruther coupled in many finite loops of granules. Many finite size loops do not produce the global phase coherence. B.G.Orr e.a. in their works has shown [7-8] that at resistance in normal state higher than RQ=*7-8 kOhm* the film have the so-called resistance reentrant behaviour. This means that our system with normal state resistance of about tens of *MOhm* (as can be seen from the Fig.1) cannot attain a zero-resistance state. Consequenly, a different technique is required to verify the superconducting transition when it does not result in a zero-resistance state. Such techinique is based on the reversed Josephson effect. The reversed Josephson effect (*RJE*) represents the inducing of the dc voltage under microvave irradiation. The induced dc voltage varies in a random, oscillatory manner as a function of temperature since it is a result of a series combination of a large number of individual junctions whose individual induced dc voltages also vary in some random oscillatory manner as a function of frequency and amplitude. At the *RJE* experiments it is important to distinguish between the Josephson *DC* voltage and rectification effects and thermal emf's. In Refs.[9-10] the self-consistent procedure for *RJE* verification has been developed. This procedure based on the study of *RJE* in confirmed high-temperature superconductors and then applied the inferences obtained to some metastable *SC* phases with the temperature well above today reached. This technique has been replicated in our work under study the electromagnetics in carbon films.

The effect of microwave detection we have observed as the occurrence on carbon film surface contacts of *DC* voltage under microwave irradiation. The value of a *DC* voltage on the detector exponentially decreased with increase in temperature of the sample and tends to be zero at ~*370 C* (see Fig.2) [6].

Experiments with carbon films were curried out in the Laboratory of superconductivity and magnetism of the Leipzig University, Germany [6,11].

Electromagnetic properties of thin *CA* films were studied by means of *SQUID* magnetometer, magnetic force (*MFM*) and electron microscope (*EM*), automatic installation on measurement of temperature dependence of electroresistance, etc. In the experiments the earlier results [1-2] on spasmodic change of electroresistance of carbon films in the critical point of voltage-current characteristic and the effect of detecting of the microwave have been confirmed. Besides the effect of magnetization oscillation of a stack of carbon films in *SQUID* magnetometer has been found out. According the *HTSC* logic each oscillation corresponds to entering of one magnetic quantum in the loop of Josephson's contacts. Also it is well known that magnetic vortices are the curriers of these magnetic quanta. So it is possible to think than the movement of magnetic vortices can explain switching of conductivity on Fig.1 and the relaxation time. The oscillations were look like to the well-known Fraunhofer picture of critical current on the Josephson contact in a magnetic field with only one difference that occurred in the much greater magnetic fields *1-3 Tesla*, than in known Josephson contacts. The difference has been explained by essentially smaller sizes of Josephson contacts in carbon films *(~100* nanometers) in comparison with the studied contacts (tens microns). The presence of so small contacts or their chains was revealed by means of *MFM* and *EM*. The topological and magnetic clusters with the sizes of *~100* nanometers has been revealed which positions were mutual correlated (see Fig.6 in the Ref.[6]). It is hoped to believe these magnetic clusters are the magnetic vortices. The dark lines of the electric currents, which were clearly visible in a vicinity of clusters by means of a *MFM*, were certified in favor of Josephson nature of the observed clusters. Preliminary magnetization of a film by a constant magnet strengthened the shielding currents.

Without dependence from a possible explanation the considered features of their behavior the carbon films can be used for creation of electronic field-effect switchers, generators and detectors of microwave radiation [12].

In the given work the results of experiments on search of possible optical radiation from a carbon film in a transitive mode of spasmodic change of electroresistance are presented. In fact the spasmodic change of conductivity in Josephson's media is connected with the breakdown of superconductivity, which should be accompanied by coherent optical radiation which frequency is unequivocally defined by the value of an energy gap of a superconductor.

## 3. Measurement of response speed of the field-effect switcher (FES) on the basis of a carbon film

*FES* carries out a role of protection against short circuits which problem in the power systems of any class always was very actual. Any protective elements or devices, whether it is safety locks, fusible inserts or switches at full conformity to normative specifications and a demanded time between failures, do not exclude the probability of the random failure. The essence of such refusals consists that during the moment of occurrence of short circuit (it is unimportant, for what reason it happens) the current strength required for melting any wires from aluminum, copper or steel avalanchely grows. This increase has the electronic speed, and the switchers having in their design the mechanical elements, always possess the sluggishness. It, in turn, leads to sticking or welding of demountable contacts and then failure, the tragedy or accident becomes inevitability. Any attempts of improvement of current switchers do not lead to absolute reliability. As a result, as electricians sadly joke, the safety lock burns down the last. The Josephson *FES* differs from the available analogues of household switches and current limiters that it is the completely electronic device in which there are no mechanical contacts, the relay, bimetallic plates and the other similar accessories, which are necessary for traditional household devices. So far the room temperature Josephson *FES* is not realised anywhere. In the *USA* the Josephson *FES* will be realised on the base of fullerenes with working temperature of *11K* [13]. Switching from a condition with high conductivity in a condition with low conductivity occurs exclusively under action of change of an electric field in the device, i.e. because of growth of a current. The current-voltage characteristic of an active element of Josephson *FES* on an initial stage of growth of a current (voltage) changes smoothly enough, as a first approximation - linearly, however in a point with critical value of a current or a voltage there is sharp strictly vertical falling of conductivity on *4-5* orders of magnitude (see Fig.1). Thus the Josephson *FES* turns to be the resistor with the resistance in tens of *MOhm*, i.e. an insulator. As it is possible to see from Fig.1 falling of a current in a critical point occurs strictly vertically, that can be the consequence of high enough speed of switching. Direct measurement of this value however is of interest.

For the measurement of switching time of *FES* the electric circuit presented on Fig.3 has been established. Here are *VS* - a voltage source, *CF* - a carbon film. Measurements were spent with carbon films of two types: *CA* and produced by chemical vapor deposition (*CVD*). The feature of the last type of films they need not to be annealed before using as it is done for *CA* - films. The electric contacts to the film surface were made by a method of vacuum deposition, and by means of an amalgam of metal gallium with copper in the ratio *50:50*. Besides were used as well the mechanical contacts. The results of measurements did not depend on type of contacts.

As the measuring device the four-channel digital oscilloscope *LeCroy WaveRunner 6100* manufactured in *USA* was used. The maximal frequency of digitization of *10 GHz* that allows to measure the processes in due course increase up to *225* picoseconds, a memory size on one channel - *2 MB*. As a source of a voltage the standard block of stabilized voltage supply *TEC-9* with a limiting voltage of *100 V* and a current of *250 mA* was used. The electric scheme of Fig.3 was constructed so that, on the one hand, to provide some voltage drop on the active element *CF*. For this purpose in condition of stable voltage on an output of a voltage source *VS* the resistor $R_1$ is entered in the scheme. Thus, at catastrophic drop of current on the active element *CF*, the voltage drop on it also will sharply decrease, that will be compensated by increase in a voltage drop at the resistor $R_1$. On the other hand it is necessary to decouple on a direct current of a circuit of the voltage supply and an oscilloscope. For this purpose the capacitor $C_1$ is entered into the scheme which capacity is chosen such that the time constant of circuit $C_1$ - oscilloscope was much greater than expected times of transients in circuit *VS* - $R_1$ - *CF*.

At measurement of parameters of transient in *CF* the circumstance is used, that at the sharp termination of a current in *CF* the potential of the left plate of capacitor $C_1$ also changes (with characteristic time of transient in circuit *CF*). That causes the recharge of capacitor and, consequently, an alternating current in a circuit of an oscilloscope. Thus, the oscilloscope, which is synchronize on single events, can define the parameters of transient, i.e. the switching time of *CF*. The results of such measurements are presented on Fig.4. As it is possible to see from Fig.4 the time of increase of transient front makes ~*1* nanosecond. As a whole the plot on Fig. 4 reminds falling off of an alternating current in a circuit with attenuation. As it is well known that the period of oscillations of a damping alternating current differs from those without attenuation a little. As shown in the work [14] the active josephson element can be considered as set of internal capacity and inductance so, that instead of Fig.3 we have received the equivalent scheme of a delay line presented on Fig.5. Internal inductance *L* and capacity $C_2$ of Josephson Element (*JE*) are rather important parameters describing its structure, geometry and electrodynamics. The opportunity of definition of values of *L* and $C_2$ through the parameters of the transient represented on Fig.4 therefore is of interest. Further we shall describe the procedure, allowing defining the specified key parameters. To define the *L* and $C_2$ we need two equations connecting these values with the data obtained during the measurement. First of such equations connects *L* and $C_2$ with the period of damping oscillations directly measured from the plot of Fig.4 is follows:

$$\frac{1}{\sqrt{LC_2}} = \frac{2\pi}{T_o}, \quad (1)$$

where $T_o = 1.2 \cdot 10^{-8}$ *sec* is the period of oscillations of an alternating current on the plot of Fig.4. The second equation can be received, having noticed, that $R_1C_2$ represents integrator, with the attenuation time of transients is $3\tau = 2.5 \cdot 10^{-8}$ *sec.*, where $\tau = R_1C_2$ is a time constant of an integrator. Whence we have obtained: $C_2 = 170$ picofarads. Then from the relation (1) we have obtained: $L_1 = 1.5 \cdot 10^{-8}$ Henri. On Fig.6 the plot of transient in the *CA* film is presented. As it is possible to see the time of increase of front of transient in this case not strongly differs from *CVD* film and makes nearby *2* nanoseconds.

The switching time between high to low conductivity at a level of *1-2* nanoseconds do not allow speaking about the thermal mechanism of switching and, more likely, testify to its electronic character. Certainly, on the base of data obtained so far it is impossible to make the unequivocal conclusion about the superconducting mechanism of switching, however it is necessary to note, that superconducting switches have the similar times of switching [15]. It is necessary to note, that the plots of transients shown on Figs.4 and 6 are not something exclusive; on the contrary, they are any way chosen from the large number of similar to them in view of time switching of the transients registered in our experiments. All this gives to hope, that superconductivity nevertheless can be the reason operating anomalous electrodynamics of some carbon condensates.

## 4. Research of processes of microwave - radiation generation in the graphite-like films.

Apparently from Fig.1, during the moment of switching power $P$ developed by a current makes some Watt. At the area of a surface of the film sample of ~$1$ $cm^2$ it can produces the specific power of a household iron at a difference of weights in $10^4$ times. During the moment of switching all power allocated on a linear path of current-voltage characteristic, should turn to heat or radiation. We shall estimate change $\Delta T$ of temperature of a film per second till the moment of switching with the well-known relation:

$$P = mC\Delta T \, . \quad (2)$$

Substituting the values of $m$~$200$ $\mu g$ which is the weight of the film sample, $C$~$1000$ $J / kg °K$ - a thermal capacity of graphite, we have obtained:

$$\Delta T = 5000\text{-}10000 \text{ °K/sec.} \quad (3)$$

Such speed of a warming up could fuse both a carbon film, and their quartz substrate. Simple estimations show, that temperature cannot be sufficiently reduced due to heat removal by heat conductivity or air convection. In a reality the temperature of a film in our experiments does not exceed $100^oC$. This can be explained only due to heat removal by self-radiation. All heated objects radiate photons with a spectrum of a black body. Distinction is that in the case of Josephson media radiation should be coherent and consist of one basic harmonic and several subharmonics. The main radiation frequency of Josephson media is defined by the critical temperature $T_c$ of superconducting transition and at $T_c$ ~ $650^oK$ makes $10^{14}Hz$, that corresponds to wavelength of an about $2$ microns in the $IR$ band. As it has been shown above, that duration of an impulse of switching makes $1$-$2$ nanoseconds then in the case of transformation of all power of a current into coherent radiation we have obtained the tempting prospect of creation of $IR$ laser with pulse power as high as $1$ $GWt$.

The simple and cheap way of registration of *IR* is the using of the photo diode. Modern photo diodes possess the good sensitivity, a suitable spectral range, and *PIN* - photo diodes have high speed - some of their modifications can process the signals with front of increase in hundreds picoseconds. However as a first step we have tried to register *IR* on a linear path of current-voltage characteristic. For this purpose the domestic photo diode *FD-7* with the wide aperture of an entrance window and a maximum sensitivity on the wavelength of *0.9* microns and a high level of internal amplification of the signal was used. This allows applying the signal directly on an oscilloscope. The limiting registered wavelength of *FD-7* is bounded by the value of *1.5* microns caused by the absorption of *IR* in a glass entrance window. Speed of the diode response does not exceed *1 millisecond*; therefore with its help it is impossible to register the pulse generation. The electric scheme presented on Fig.7 has been established for the measurements.

The results of measurement are presented on Fig.8. On Fig.8 it is possible to see a thermal impulse with duration of *3* milliseconds on which background the high-frequency generation is visible. It is known [15], when biasing with *DC* voltage *V* the josephson's contact starts to generate the radiation with the frequency *v/V=483.6 MHz/µV*. Using this relation and the data of Fig.1, it is possible to estimate the number of Josephson's contacts as high as *~300*. From Fig.8 it is visible, that in the field of a maximum of thermal peak the amplitudes of generating pulses considerably increase, that, possibly, is connected with increasing the instability of an electrodynamics condition. Apparently, high-amplitude peaks in area of a maximum of thermal peak reflect the energy release by means of generation of short microwave pulses, which cannot be fully resolved by photo diode *FD-7* because of its low response speed.

The given conclusion proves to be true that in a point *0.5 ms* on the plot of Fig.8 sharp decrease of a thermal pulse is visible that, at least, can be partially caused by the microwave radiation. The basic peak of the microwave radiation should be characterized by width in some nanoseconds, which is caused by time of switching from highly conductive state (see Fig.4 and Fig.6) and cannot be registered by means of the slow photo diode.

## 5. Registration of IR radiation with the fast photodiode

For registration of *IR* radiations during the moment of switching the fast silicon avalanche photo diode have been used which is applied in fibre-optical lines of communication. This device posses the response speed in hundreds of picoseconds and a spectrum of registration from *600* nanometers up to *1500* nanometers. The working window of the photo diode in diameter of *1.1 mm* consists of *556* pixels that enable to register radiation in a wide angular cone. Sensitivity of the photo diode makes *0.8 mV/photoelectron*, and the photoelectron is born in *90* % of cases of interaction of photons with sensitive substance of the photo diode. The scheme of inclusion of the photo diode is presented on Fig.9. As the photo diode is sensitive to visible light the registration of radiation from a carbon film near to the moment of switching of conductivity was made in the blacked out chamber. The sensitive area of the photo diode settled down near to a radiating surface of a carbon film. The oscilloscope was forced to work in a mode of synchronization from single pulses. The photo diode was biased with *39.4 V* of *DC* voltage and its reaction to visible light was checked. At the closed cover of the blacked out chamber the background indications of an oscilloscope did not exceed *1 mV*. Then the *DC* voltage on contacts of a carbon film slowly increased up to critical value at which the current in a circuit of a film jump fell up to zero value. During the moment close to switching an oscilloscope registered the optical radiation, the plot of which time dependence is presented on Fig.10. On a background of "substrate" it is visible a series of optical pulses which amplitudes considerably exceeds a level of "substrate" so enters the photo diode into a condition of saturation. Division of process of generation into some stages can cause the presence of several consecutive optical impulses. Such representation contradicts with the data on much more transient and coherent switching of conductivity during *1* nanosecond. However as it was mentioned before the oscolloscope was synchronized on single switching impulse. So it is possible to suppose that each optical impulse corresponds to switching one. That is in reality in the case of optical radiation of Fig.10 the true switching picture consist of corresponding number of swithing pulses, from which the oscilloscope visualized onle the first. The fact that the duration of optical pulses considerably exceeds the switching time may be explained due to limited propagation speed of optical radiation in carbon film body. Time of propagation of radiation in a carbon film with the thickness $h\sim1\mu m$ till the moment of switching will make $\tau\sim h^2C/\lambda \sim 5\cdot10^{-6}$ sec. The delay can increase due to decrease of heat conductivity $\lambda$, which is connected with conductivity $\sigma$ by the well-known Wideman –France law:

$$\lambda = L_0\sigma T, \qquad (4)$$

where $L_0 = 2.445\cdot10^8$ $WOhm/^0K^2$- is the Lorentz's number, $\sigma$- the electrical conductivity of a film, $T$ - the temperature. As can be seen from Fig.1, the conductivity of a film before switching makes $2\cdot10^4$ $(Ohmm)^{-1}$. If to assume, that after switching the conductivity will fall in $10^4$ times as on Fig.1 than $\tau$ will make even *50 ms*. Of course there is no reason to consider the all parts of film as insulator. So the time delay can be successfuly explained.

In fact, as it was mentioned before, in the absence of global phase coherence the conductivity of granular system has the percollation character. This means the system is composed by number of branches of finite size conduction clusters. So the swithing process may consist from a few steps of consequent "turning off" of individual branches of conductive clusters. The first turn off happens in the branch where the current first exceeds the critical value. Then the remaining branches turn off consequently resulting at last in zero current. It is believed this picture is broadly in line with the data of Fig.10. If the motion of magnetic vortices explains the change of conductivity in carbon films, than Fig.10 displays the transformation of vortices into the *IR* radiation. This gives us some idea of new type of laser based on the motion of magnetic vortices. In this case we can consider the Josephson media of granular structure as laser active media. Due to pinning of vortices in the film body their movement is thermaly activated. That is the long relaxation time is needed for the recovery of conductivity after switching. Hovewer it is expected that under intense *IR* irradiation the relaxation time will be decreased like to pumping process in laser.

As to the wavelength of optical radiation from a carbon film it is possible to tell, that, first, it lies outside of an interval of visible light. Secondly, this length of a wave obviously is less than *1.5* microns - the limit of transmission for the silicon photo diode. Thus, it is possible to define with a high degree of probability, that the wavelength of radiation lays near to *1* micron that agree well with the mechanism of radiation connected with the destruction of magnetic vortices in a superconductor with the critical temperature *650-700K*.

## 6. Acknowlegements

We thank Yu.G.Kudenko for granting of the fast photodiode manufactured by CPTA (Moscow) for study of generation of optical radiation.
Authors are grateful to the Russian Foundation of Basic Researches for support this work by the grant №05-08-17909-a.

**Figure captions**

Fig.1 Current-voltage characteristic of *FES* based on *CA* film. Two variants of switching are represented with manual and authomatic current feeding on the sample. Switching time is about *1* nanosecond. The ratio of conductivities before and after switching is about $10^4$-$10^5$.

Fig.2. Temperature dependence of the *DC* voltage induced by microwave radiation. The voltage start to be zero at *650K*.

Fig.3. The electric scheme of measurement of switching speed of *FES*.

Fig.4. The time dependence of transient current in carbon *CVD* film. The swithing time is defined as the time to first maximum of current.

Fig.5. The equivalent electric scheme for measurement of transient parameters under switching.

Fig.6. The time dependence of transient current in *CA* film. Again like Fig.4 the swithing time is defined by a point of first extremum.

Fig.7. The scheme of *IR* registration on the linear path of current-voltage characteristic of a carbon film.

Fig. 8 Results of registration of radiation from *CVD* carbon film in a stationary mode by means of photo diode *FD-7*.
Fig.9. The scheme of connection of the high-speed photo diode.
Fig.10. The optical radiation registered by the fast photo diode from a *CVD*-carbon film near to the moment of switching.

**References**

1. S.G.Lebedev and S.V.Topalov, "*Evidence of Weak Superconductivity in Carbon Films*", **Bulletin of Lebedev's Physical Institute**, N12 (1994) 14-20.
**2.** S.G.Lebedev, "*Anomalous Electromagnetics of Carbon Condensates in the Light of Ideas of a High-Temperature Superconductivity*", **arXiv: cond-mat/**0510304.
3. K.Antonowicz e.a. **Carbon**, 11 (1973) 1.
4. K.Antonowicz e.a. **Carbon**, 10 (1972) 81.
5. K. Antonowicz, **Nature, 247**, 358 – 360, 1974.
6. S.G.Lebedev "*Particle Irradiation for Verification of Superconducting-Like Behavior in Carbon Arc Films*". **Nucl. Instr. Meth.** A521 (2004) 22-29.
7. B.G.Orr, H. M. Jaeger, and A. M. Goldman, **Phys.Rev**. **B32**, 7586 (1985).
8. H.M.Jaeger, D. B. Haviland, A. M. Goldman and B. G. Orr, **Phys. Rev**. **B34**, 4920 (1986).
9. J.T.Chen, L. E. Wenger, C. J. McEwan and E. M. Logothetis, Phys**. Rev. Lett. 58,** 1972 (1987).
10. R. Munger and H. J. T. Smith, **Phys. Rev**. **B44**, 242 (1991).
11. S.G.Lebedev, "*Exploration of Strange Electromagnetics in Carbon Films*". **arXiv: cond-mat/**0509535.
12. S.G.Lebedev, **Patent of Russian Federation** N 2212735 on Sept. 20, 2003.
13. J. H. Schon, Ch. Kloc, R. C. Haddon, B. Batlogg , **Science 288**, 656 (2000).
14. R.Parmentie, In Solitons in Action (ed. K.Longren and A.Scott), Academic, 1978, p.184-189.
15. A.Barone, G.Paterno, "Physics and Applications of the Josephson Effect" (Wiley, 1982).

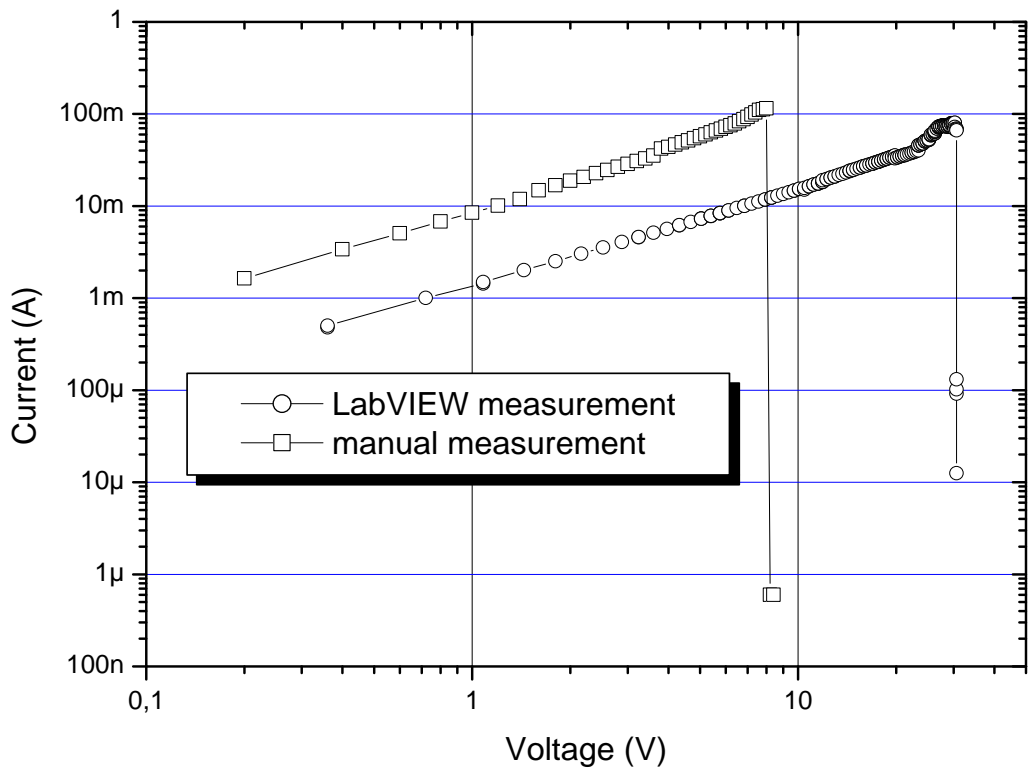

Fig. 1

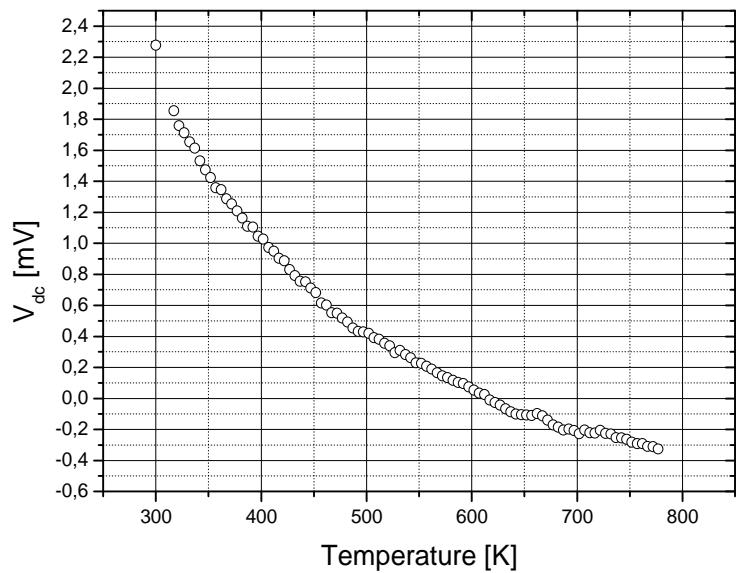

Fig.2

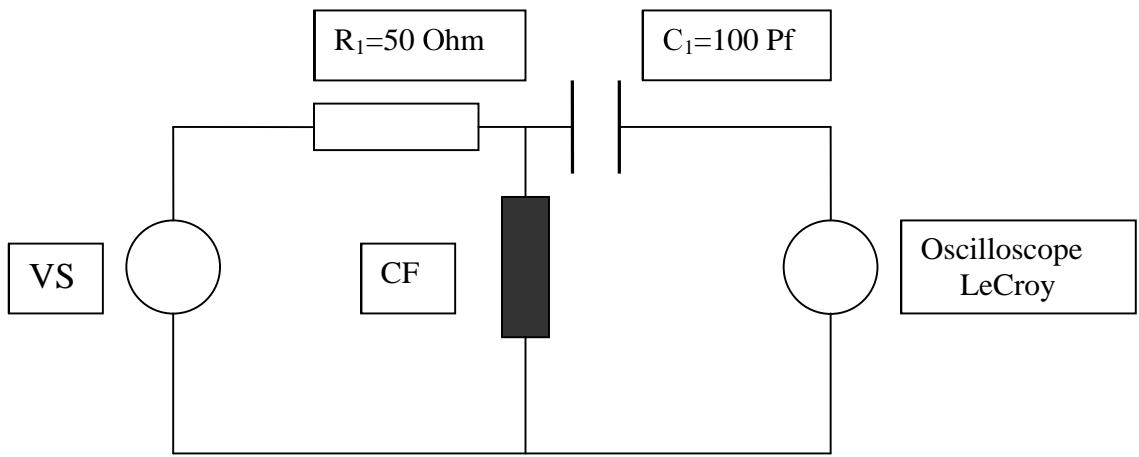

Fig.3

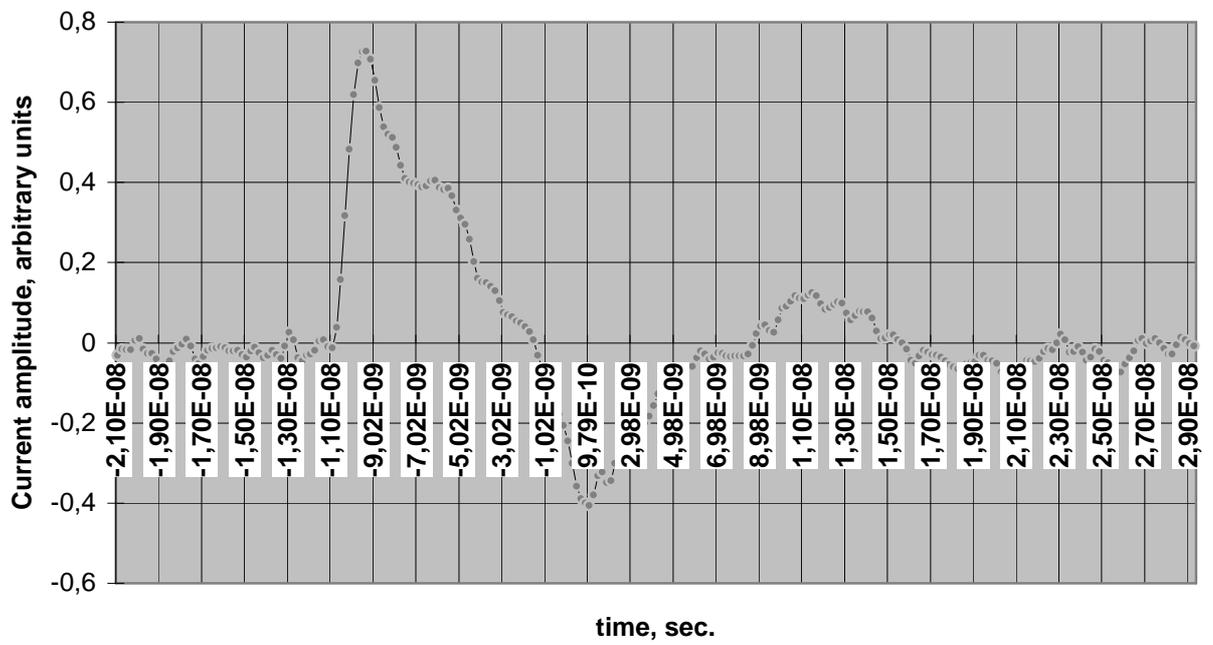

Fig.4

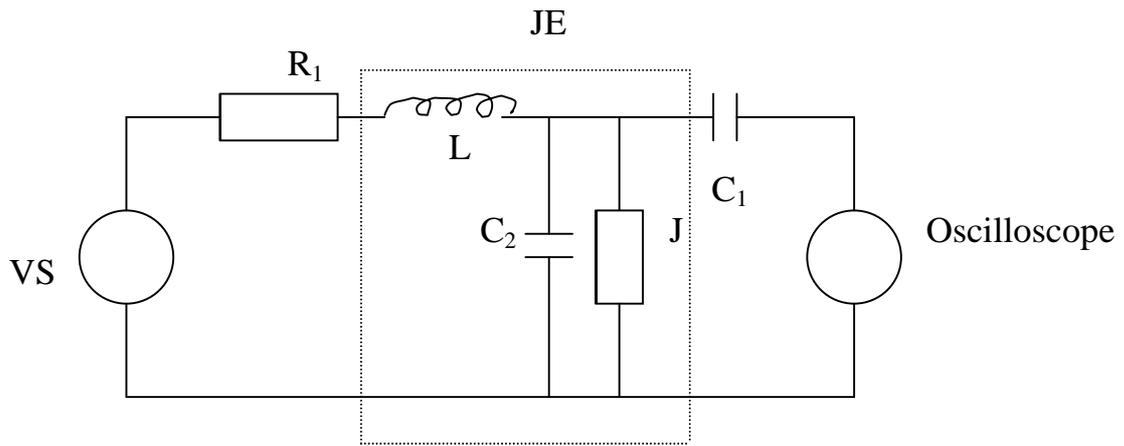

Fig.5

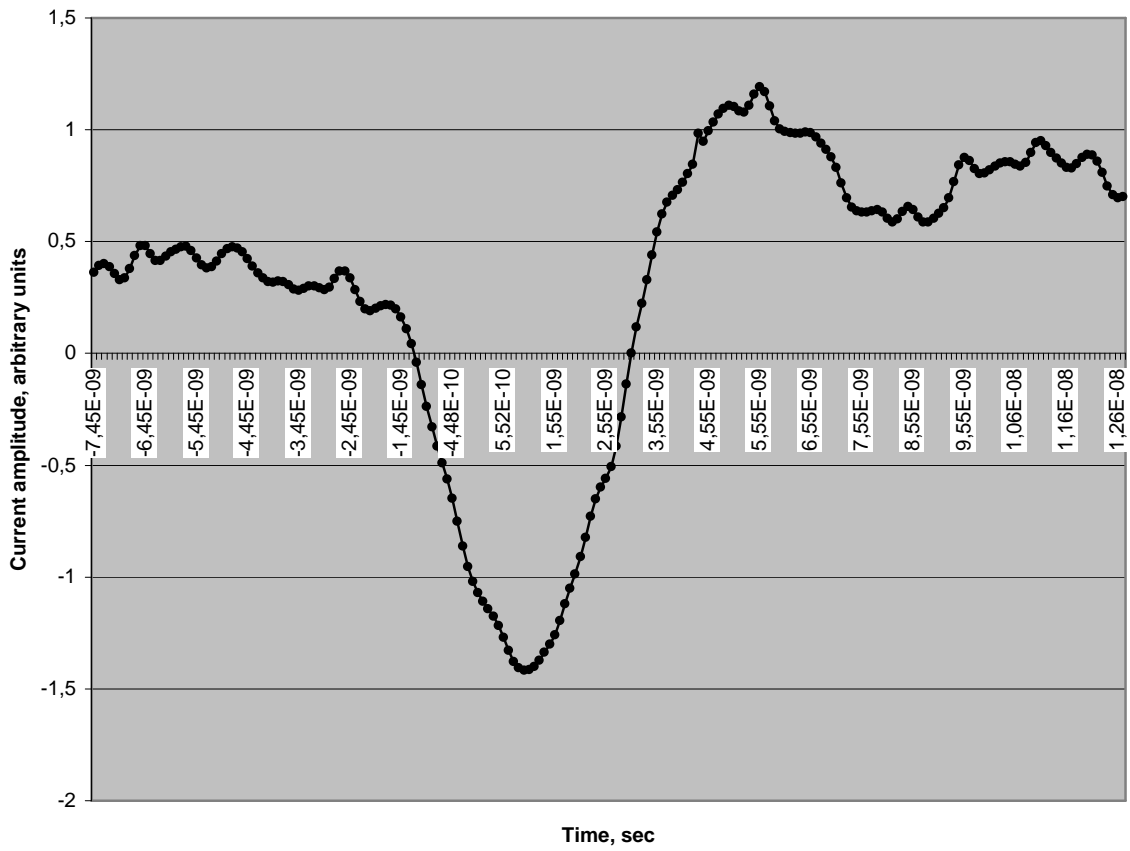

Fig.6

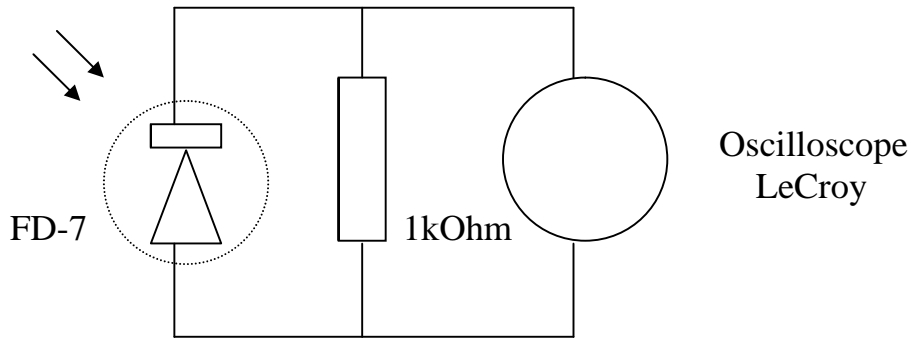

Fig.7

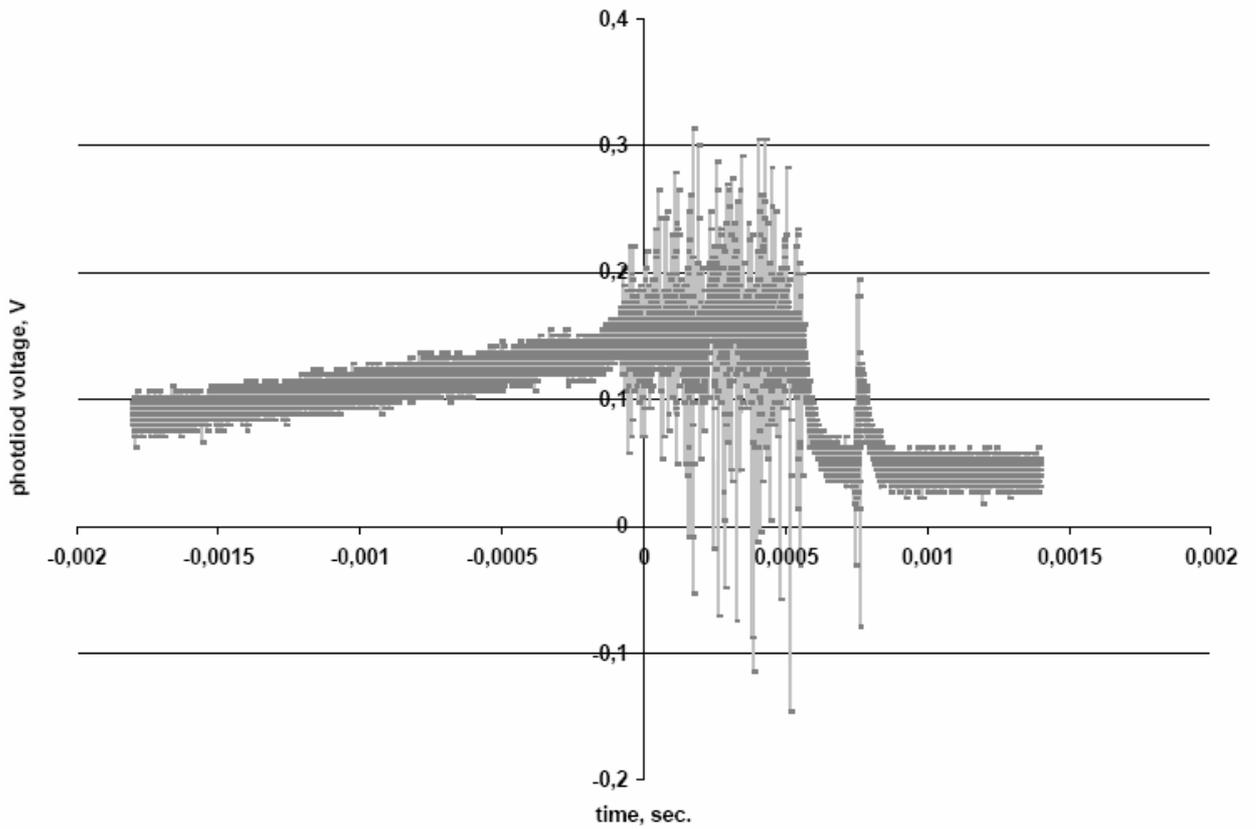

Fig.8

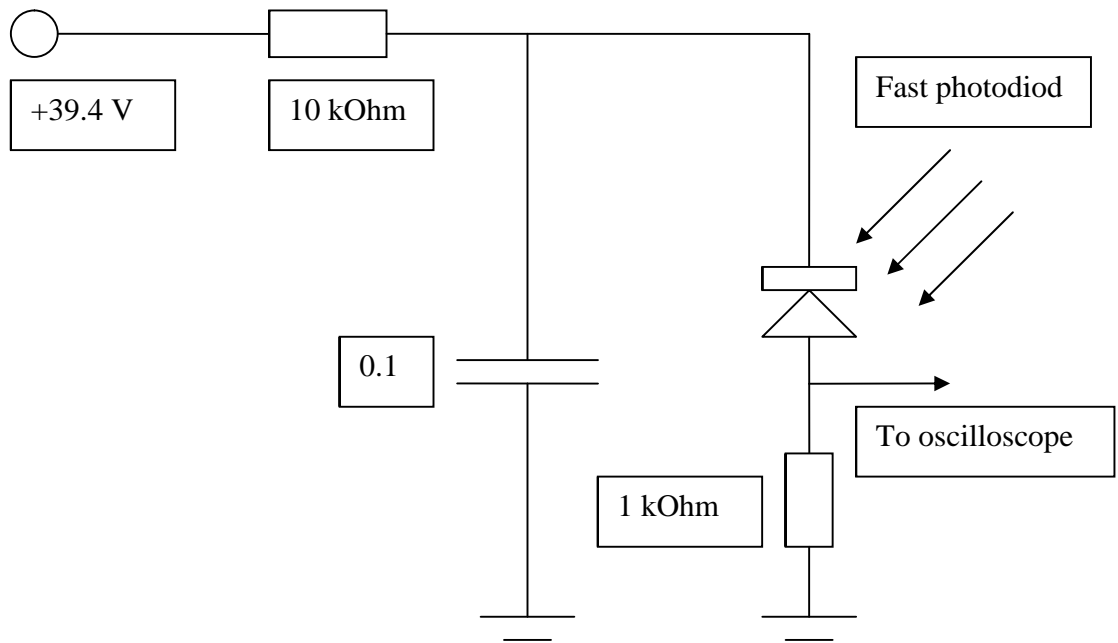

Fig.9

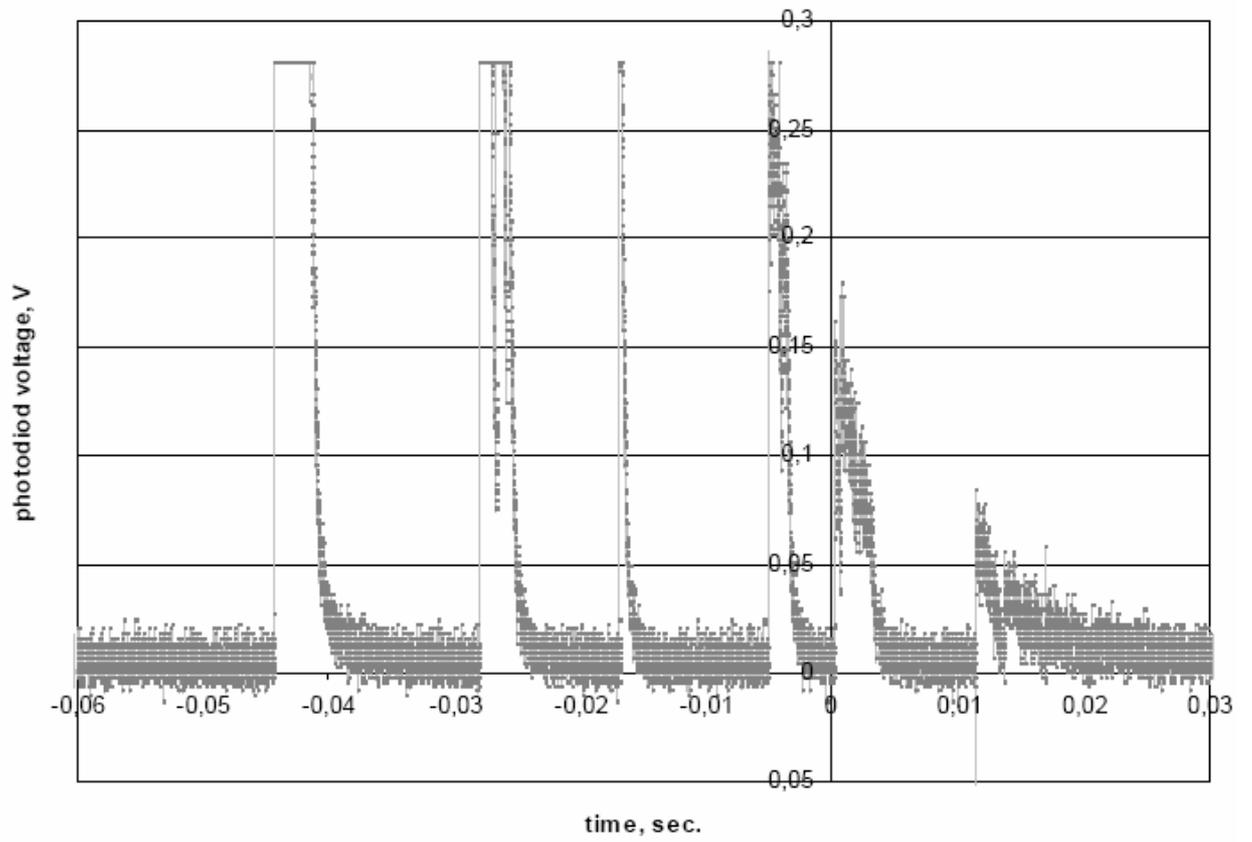

Fig.10